\documentclass[prb,superscriptaddress,twocolumn]{revtex4}
\usepackage{epsfig}
\begin{document}
\title{Electron spin and charge switching in a coupled quantum dot quantum ring system}
\author{B. Szafran}
\affiliation{Departement Natuurkunde, Universiteit Antwerpen
(Campus Drie Eiken), B-2610 Antwerpen, Belgium}
\affiliation{Faculty of Physics and Nuclear Techniques, AGH
University of Science and Technology, al. Mickiewicza 30, 30-059
Krak\'ow, Poland}
\author{F.M. Peeters}
\affiliation{Departement Natuurkunde, Universiteit Antwerpen
(Campus Drie Eiken), B-2610 Antwerpen, Belgium}
 \author{S. Bednarek}
 \affiliation{Faculty of Physics and Nuclear Techniques, AGH
University of Science and Technology, al. Mickiewicza 30, 30-059
Krak\'ow, Poland}

\date{\today}

\begin{abstract}
Few-electron systems confined in a quantum dot laterally coupled
to a surrounding quantum ring in the presence of an external
magnetic field are studied by exact diagonalization. The
distribution of electrons between the dot and the ring is
influenced by the relative strength of the dot and ring
confinement, the gate voltage and the magnetic field which induces
transitions of electrons between the two parts of the system.
These transitions are accompanied by changes in the periodicity of
the Aharonov-Bohm oscillations of the ground-state angular
momentum. The singlet-triplet splitting for a two electron system
with one electron confined in the dot and the other in the ring
exhibits piecewise linear dependence on the external field due to
the Aharonov-Bohm effect for the ring-confined electron, in
contrast to smooth oscillatory dependence of the exchange energy
for laterally coupled dots in the side-by-side geometry.
\end{abstract}
 \maketitle
\section{introduction}

Coupling\cite{v2,v3,v4,v5,v6,l1,l2,l3,le1,le2,le3,GB,Wens,Harju}
between semiconductor quantum dots\cite{hawrylak} results in the
formation of so-called artificial molecules. Since most of the
quantum dots have flat geometry, the coupling is realized either
by vertical stacking\cite{v2,v3,v4,v5,v6} or by fabrication of
dots coupled laterally on the same
plane.\cite{l1,l2,l3,le1,le2,le3,GB,Wens,Harju}
Theoretical\cite{l1,l2,l3,GB,Wens,Harju} considerations and
experimental\cite{le1,le2,le3} realizations of laterally coupled
dots are based on the idea of dots placed side by side. This paper
is devoted to few-electron states in an essentially different
geometry of lateral coupling, namely, to a quantum dot surrounded
by a quantum ring\cite{rev} with a tunnel barrier separating both
parts of the system. The confinement potential considered in this
paper can be obtained using an atomic force microscope (AFM) to
locally oxidize\cite{oxt} the sample surface which results in the
depletion of the two-dimensional electron gas (2DEG) underneath
it.  Alternatively one can apply split gates with a central cap
gate surrounded by a thin collar gate on top of a planar
nAlGaAs-GaAs heterostructure containing a 2DEG. A proper geometry
of split gates for the fabrication of the confinement potential
considered in this paper was applied in the study\cite{ZN} of
effects related to electron localization on local fluctuations of
the confinement potential in the low electron density regime. The
system studied in the present paper would require a sufficiently
strong confinement which is less perturbed by fluctuations. The
effect of local perturbations can be largely diminished by
optimization\cite{Lis} of the size of electrodes for the strength
of the electrostatic confinement potential.

Phase effects appearing in electron transport through quantum dots
were studied in the Aharonov-Bohm interferometer.\cite{K1,Y1} The
potential geometry studied in this paper is a two-dimensional
counterpart of quantum-dot quantum-well
structures.\cite{qdqw1,qdqw2} Impurity effect on the
single-electron states in a three-dimensional quantum ring for
strong in-plane confinement has been studied.\cite{badziew}
Related to the present work is the magnetic coupling of a
superconducting disk surrounded by a superconducting
ring.\cite{baelus} In contrast to the work of Ref. [\cite{baelus}]
in the system considered here the coupling between the ring and
the dot occurs through quantum mechanical tunnelling.

We study the effect of the magnetic field on the confined one, two
and three-electron systems using an exact diagonalization
approach. In quantum dots and rings the magnetic field induces
ground-state angular momentum transitions. However, the role of
the electron-electron interaction for the transitions in these two
structures is different. In quantum rings the interaction is of
secondary\cite{CP,CEPL} importance for the angular momentum
transitions which are mainly determined by the Aharonov-Bohm
effect. In spinless\cite{CP,RM} few-electron systems the
ground-state angular momentum is not influenced by the Coulomb
interaction and for electrons with spin the angular momentum of
the ground state differs from the noninteracting case by at most
$\hbar$.\cite{CEPL} On the other hand, in quantum dots the Coulomb
interaction influences strongly the values of the magnetic field
at which the angular momentum transitions appear. Moreover, in
two- and three-electron systems these transitions are absent if
there is no electron-electron interaction. In this paper we study
the hybrid magnetic-field evolution of the electron spectra in the
dot-in-the-ring geometry.

 The magnetic-field along with the angular momentum
transitions induces a redistribution of the electron charge in
quantum dots.\cite{RM,SBA} Here, we will show that in the
considered geometry the magnetic field can be used to transfer the
electrons from the dot to the ring or {\it vice versa}. We will
also address the problem of the magnetic-field-induced trapping of
electrons in local potential cavities.\cite{bunching}

The spins of a pair of electrons localized in laterally coupled
dots have been proposed\cite{GB} as a possible realization of
coupled qubits. A universal quantum gate requires the possibility
of application of single qubit as well as two-qubit rotations. For
this purpose one should be able to address each of the electrons
individually as well as to control the state of the pair, which
requires the spatial separation of electrons and a tunable
coupling between them. We studied the singlet-triplet splitting
energy for the two-electron system with one electron localized in
the dot and the other in the ring. We show that the angular
momentum transitions, appearing for the ring-confined electrons as
a consequence of the Aharonov-Bohm effect, lead to a simple
piecewise linear dependence of the exchange energy on the external
magnetic field. Since the unitary evolution in quantum computation
needs precise control of the underlying qubit interaction this
simple dependence makes our system a good candidate for the
realization of the magnetic field controllable pair of spin
qubits. Recently, it has been established\cite{hanson} that the
spin relaxation time in quantum dots defined by electric gates in
two-dimensional electron gas is much longer than the qubit redout
time in spin-to-charge conversion technique.

This papers is organized as follows: in the next section the
present approach is explained, the single electron spectrum is
described in Section III, the results for two and three electrons
are given in Sections IV and V, respectively, and Section VI
contains the summary and conclusions.

\section{Theory}
We study two-dimensional $N$-electron systems confined in circular
potentials using the effective mass Hamiltonian
\begin{equation}
H=\sum_{i=1}^N h_i + \sum_{i=1}^N\sum_{j=i+1}^N
\frac{e^2}{4\pi\varepsilon\varepsilon_0 r_{ij}}+BS_zg^*\mu_B,
\label{he}
\end{equation}
where $\varepsilon$ is the dielectric constant, $g^*$ -- the
effective Land\'e factor, $\mu_B$ -- the Bohr magneton, $S_z$ --
the $z$ component of the total spin, $B$ -- the magnetic field,
and $h_i$ stands for the single-electron Hamiltonian, which
written in the symmetric gauge  $\mathbf{A}=(-By/2,Bx/2,0)$ has
the form
\begin{equation}
h=-\frac{\hbar^2}
{2m^*}\nabla^2+\frac{1}{8}m^*\omega_c^2\rho^2+\frac{1}{2}\hbar
\omega_cl_z +V(\rho), \label{H1}
\end{equation}
with $m^*$ -- the electron effective mass, $l_z$ -- z-component
angular momentum operator,  $\omega_c=eB/m^*$ -- the cyclotron
frequency and $V(\rho)$ -- the confinement potential. We adopt
material parameters for GaAs, i.e., $m^*/m_0=0.067$,
$\epsilon=12.9$, and $g^*=-0.44$. The last term of Eq. (1), i.e.,
the spin Zeeman splitting energy is independent of the
distribution of electrons between the different parts of the
system as well as of the Coulomb interaction energy. Moreover, the
value of the $g^*$ factor can be tuned by the admixtures of Al
substituting Ga.\cite{gtune} We have therefore decided to neglect
the Zeeman effect in most of the results presented in this paper
(unless explicitly stated otherwise).

\begin{figure}[htbp]{\epsfxsize=60mm
                \epsfbox[16 120 563 650]{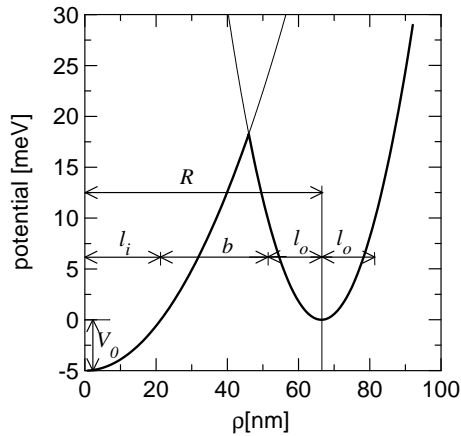}}\newline
%c:\luty\nci2d_latcaoup_harmonic\publishable\TBP\1e\fig1
\caption{Confinement potential (cf. Eq. \ref{potential}) for
$\hbar \omega_i=5$ meV, $\hbar \omega_o=10$ meV, $V_0=-5$ meV,
$b=30$ nm and the GaAs effective mass $m^*/m_0=0.067$. The dot
oscillator length $l_i=\sqrt{2\hbar/m\omega_i}$ is equal to
$21.33$ nm and the oscillator length for the ring $l_o=15.08$ nm
which gives the ring radius $R=66.4$ nm.
 \label{potentialrys}
 }
\end{figure}

 We model a strictly two-dimensional cylindrically symmetric
potential of a quantum dot placed within the quantum ring  with
the following confinement potential
\begin{equation}
V(\rho)=\min(m^*\omega_i^2\rho^2/2+V_0,m^*\omega_o^2(\rho-R)^2/2),
\label{potential}
\end{equation}
where  $\hbar \omega_i$ and $\hbar \omega_o$ are the confinement
energies of the dot and the ring respectively, and the radius of
the ring $R$ is determined by the sum of oscillator lengths for
the dot and ring potential and the barrier thickness ($b$)
according to formula
$R=\sqrt{2\hbar/m\omega_i}+\sqrt{2\hbar/m\omega_o}+b$. This
potential is parabolic within both the quantum dot and the quantum
ring, $V_0$ is the depth of the dot confinement with respect to
the bottom of the quantum ring potential. The confinement
potential (\ref{potential}) is shown in Fig. \ref{potentialrys}
for $\hbar \omega_i=5$ meV, $\hbar \omega_o=10$ meV, $V_0=-5$ meV
and $b=30$ nm.  A model potential parametrized similarly to Eq.
(\ref{potential}) was used previously for the
description\cite{Wens,Harju} of side by side quantum dots. The
cusp present in simple potentials of this type (cf. Fig.
\ref{potentialrys}) is rather unphysical and cannot be realized in
real structures, however this shortcoming is of secondary
importance since the cusp appears in a region of space where the
barrier potential is largest and the wave functions of the lowest
energy levels are small. In the weak coupling limit (for large
barrier thickness) approximate formulas for the dot- and ring-
confined states as functions of the magnetic field can be given
(see below).

In the present paper the single-electron eigenfunctions for
Hamiltonian (\ref{H1}) and definite angular momentum are obtained
numerically on a radial mesh of points using the finite-difference
approach. Eigenstates of the two- and three-electron Hamiltonian
(\ref{he}) are calculated using the standard configuration
interaction method\cite{bs1} with a basis composed of Slater
determinants built with single-electron wave functions. The
Coulomb matrix elements are evaluated by a
two-dimensional\cite{CP} numerical integration. The few-electron
states are described by the total spin $S$ and angular momentum
$L$ quantum numbers. In this paper we discuss only the two- and
three- electron states with nonpositive total angular momenta. We
will therefore omit the minus sign for the angular momentum
quantum number $L$.

\section{single-electron states}
\begin{figure*}[htbp]{\epsfxsize=180mm
                \epsfbox[33 640 460 777]{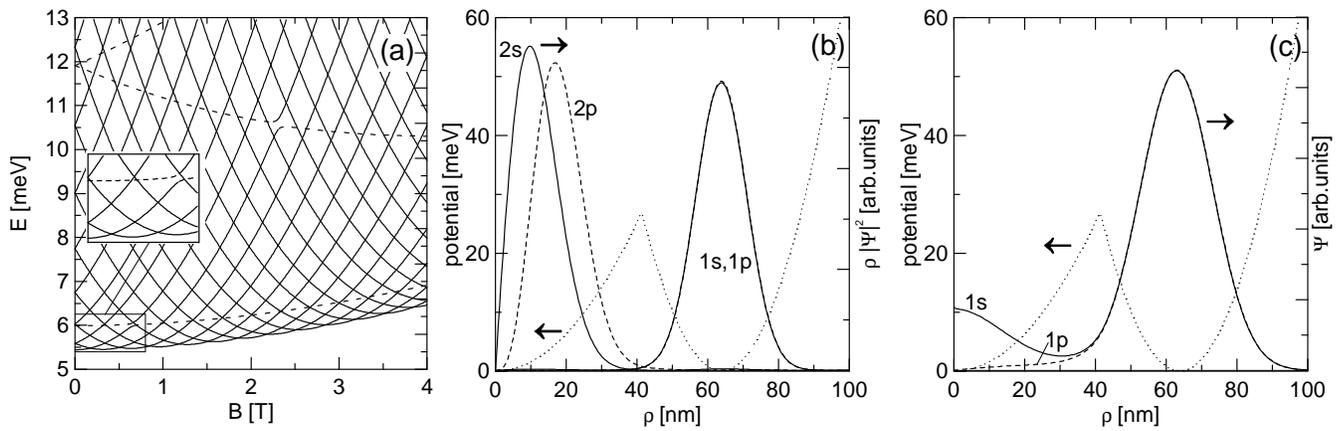}}\newline
%c:\luty\nci2d_latcaoup_harmonic\publishable\TBP\1e\fig1
\caption{(a) Single-electron spectrum for $\hbar\omega_i=6$ meV,
$\hbar \omega_o$=11 meV, $V_0$=0, and $d$=30 nm ($R=63.85$ nm).
The solid lines correspond to states localized in the ring and the
dashed lines to states localized in the dot. Lowest of the dashed
lines corresponds to $s$ state and the two higher to $p$ states.
Inset shows the low-field and low-energy part of the spectrum -
enlargement of the fragment surrounded by thin solid lines
corresponding to anticrossing of 0 angular momentum dot- and ring-
confined energy levels. Dotted line in (b) and (c) shows the
confinement potential (left scale) for the parameters applied in
(a). Solid and dashed curves in (b) show the radial probability
density $\rho |\psi|^2$ and in (c) the wave functions of the
lowest states of $s$ and $p$ symmetry, respectively.
 \label{ses} \label{fig1}
 }
\end{figure*}

\begin{figure}[htbp]{\epsfxsize=70mm
                \epsfbox[33 125 548 659]{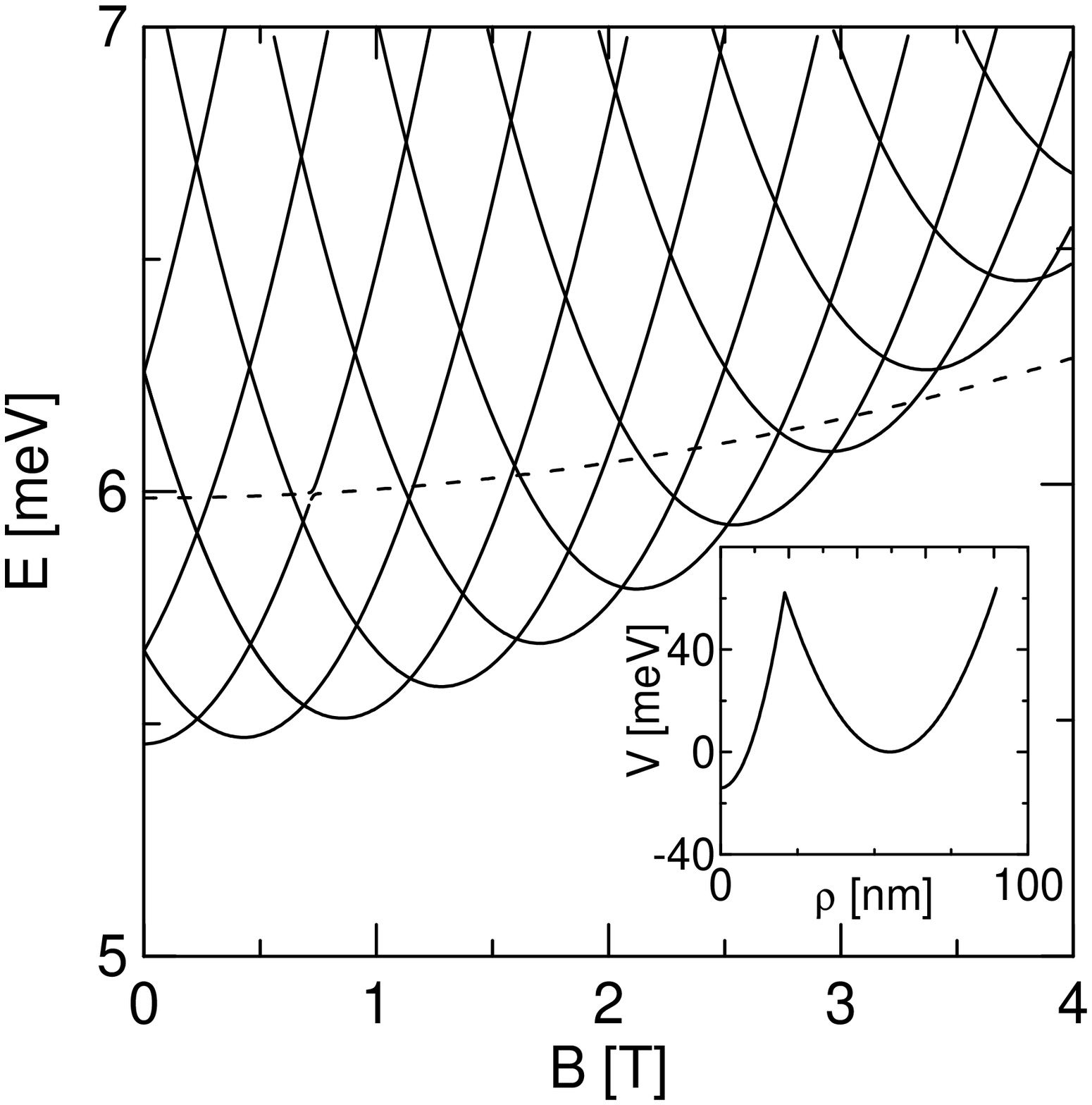}}\newline
\caption{Single-electron spectrum for $\hbar\omega_i=20$ meV,
$\hbar \omega_o$=11 meV, $V_0=-$14 meV, and $d=30$ nm (potential
is plotted in the inset). The solid lines correspond to states
localized in the ring and the dashed line to the lowest-energy
state localized in the dot.}\label{diper1e}
\end{figure}

The single-electron spectrum for $\hbar\omega_i=6$ meV, $\hbar
\omega_o=11$ meV, $V_0=0$ and $b=30$ nm is shown in Fig.
\ref{ses}(a). For this, relatively large barrier thickness, the
low part of the energy spectrum is essentially a sum of the
spectra of an electron localized in the dot and in the ring. The
solid lines in Fig. \ref{ses} correspond to states localized in
the ring and dashed lines to $s$ (lowest dashed line) and $p$
states localized in the dot. The ring part of the spectrum
exhibits Aharonov-Bohm oscillations. The angular momentum of the
lowest-energy ring-localized states takes on the subsequent values
0,-1,-2, etc. [in $\hbar$ units] when the magnetic field
increases. The period of these oscillations is 0.337 T. This
period corresponds to a flux quantum passing through a strictly
one-dimensional ring of radius $R_{1D}=62.51$ nm which is in good
agreement with the radius of the ring in the present model
$R=63.85$ nm. The energy of the states localized in the dot change
with the magnetic field more slowly than the energy of the
ring-localized states. States with the same angular momentum
change their order in anticrossings due to quantum mechanical
tunnel coupling between the dot and the ring. Anticrossing for $s$
states appears for $B$ around 0.65 T [see inset of Fig. 2(a)]. A
much wider anticrossing for $p$ states is visible around 2.4 T.

Fig. \ref{ses}(b) shows the confinement potential for the
parameters applied in Fig. \ref{ses}(a) as well as  the radial
probability densities for the lowest $s$- and $p$- symmetry
states. The radial densities for the ring-localized states do not
depend on the angular momentum. However, Fig. \ref{ses}(c) shows
that the $s$-wave function penetrates the dot region in a much
stronger way than the $p$-type wave function. It will have an
important consequence for the singlet-triplet splitting of the
two-electron states (see the next Section). Note that the angular
momentum has an opposite effect on the strength of the tunnelling
of the dot-localized states to the ring part of the potential.
Barrier thickness is effectively smaller for the dot-localized
states of higher angular momentum [cf. Fig. \ref{ses}(b)].

The dependence of the energy of the lowest dot-localized state can
be very well approximated by the expression for the lowest
Fock-Darwin state, i.e.,
$E_{dot}=V_0+\sqrt{(\hbar\omega_i)^2+(\hbar \omega_c/2)^2}$.
Without the magnetic field the lowest energy ring-localized level
is approximately equal to $\hbar \omega_o/2$, i.e., to the energy
of the single-dimensional harmonic oscillator in the radial
direction. In the external field the envelope of the lowest-energy
ring-localized level can be quite well approximated by
$E_{ring}=\sqrt{(\hbar\omega_o/2)^2+(\hbar \omega_c/2)^2}$. These
two formulas can be used in order to roughly determine whether the
ground state of a single electron is localized in the dot or in
the ring. For equal depth of the ring and the dot ($V_0=0$) the
magnetic field does not change the order of the lowest-energy dot-
and ring- confined states. However, for $V_0=0$ and
$\omega_o\approx 2\omega_i$ the magnetic field can induce
oscillations of the ground state localization from the dot to the
ring, which results from the local deviations of the lowest
ring-confined energy level from the smooth envelope [cf. Fig.
\ref{ses}(a)]. On the other hand, the magnetic field favors
localization in the deep but small (thin) quantum dot (ring). This
effect is illustrated in the following figure.

Fig. \ref{diper1e} shows the energy spectrum for a $\hbar
\omega_i$ which is increased with respect to Fig. \ref{ses} from 6
to 20 meV  and the bottom lowered by $V_0=-14$ meV. For $B=0$ the
low-energy part of the spectrum is the same as in the case shown
in Fig. \ref{ses}(a). However, the energy of the dot-localized
state grows more slowly than the envelope of the ring-localized
states. In consequence, the dot-localized state becomes the ground
state for $B=3.3$ T. When the radius of the Landau orbit becomes
smaller than the size of the local potential cavity, the electron
can enter inside the dot without an extra increase of the kinetic
energy due to the localization. Similar effects of trapping of
electrons in local potential cavities at high magnetic fields are
probably at the origin of the bunching of the charging lines
observed in single-electron capacitance spectroscopy of large
quantum dots.\cite{bunching} The opposite effect, i.e., the change
of the ground-state localization from the dot to the ring under
influence of the external magnetic field is also possible if the
ring is thin but with a bottom deeper than the dot.

\section{Two electron system}

\begin{figure}[htbp]{\epsfxsize=60mm
                \epsfbox[128 60 496 781]{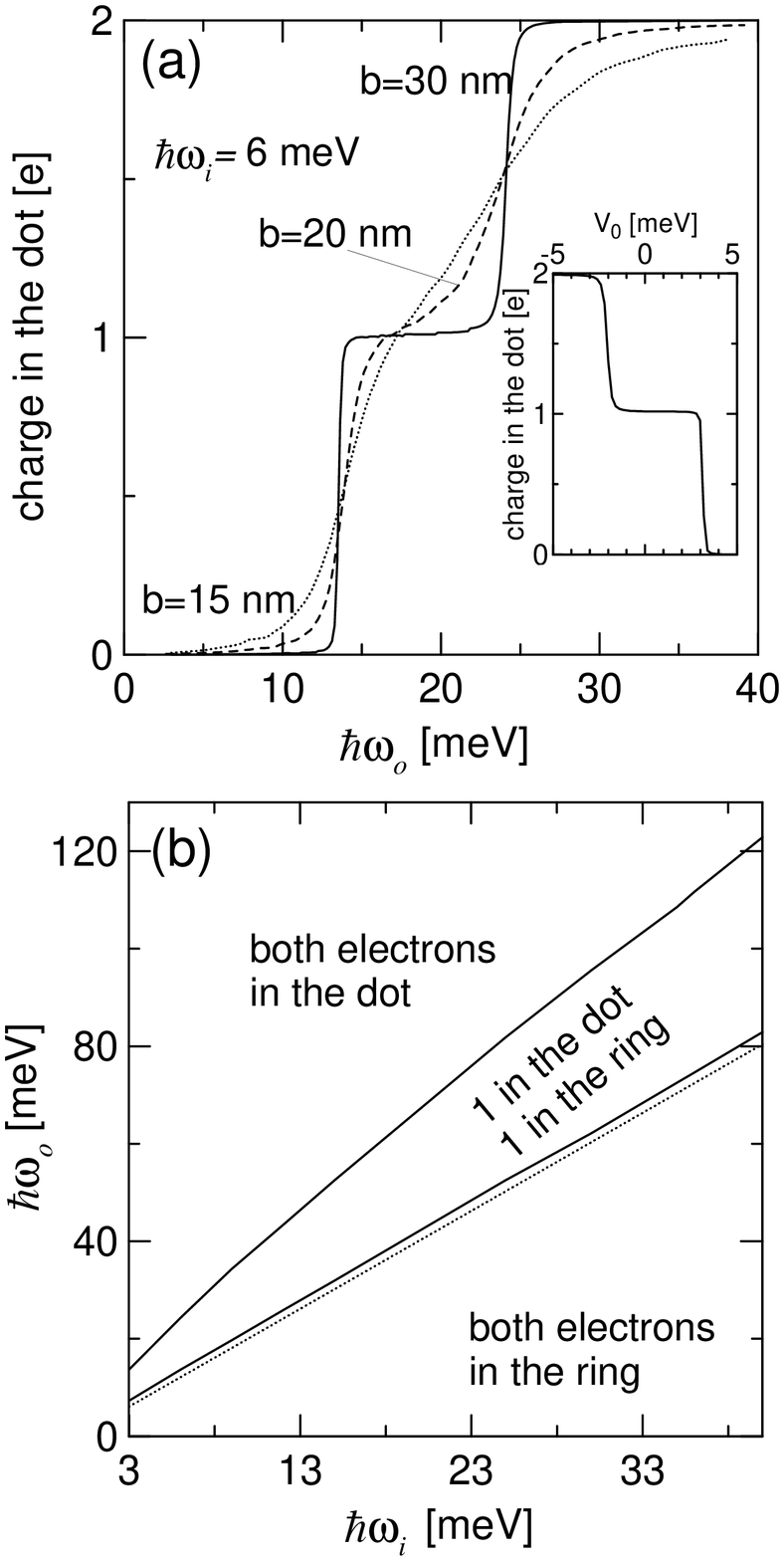}}\newline
%c:\luty\nci2d_latcaoup_harmonic\publishable\TBP\2e\fig2>
\caption{ (a) Charge accumulated in the dot as function of $\hbar
\omega_o$ for different values of the barrier thickness and
$V_0$=0. Inset in (a) shows the charge accumulated in the dot as
function of $V_0$ for $b=30$ nm, $\hbar \omega_i$=6 meV and $\hbar
\omega_o$=20 meV. (b) Phase diagram for the distribution of two
electrons for $B=0$, $V_0$=0 and $b=30$ nm.
 Solid lines in (b)
divide regions of different electron localization in the
two-electron system. Above the dotted line the ground state of a
single electron is localized within the dot. \label{fig2}}
\end{figure}

For $B=0$ the ground state of the electron pair corresponds to
zero total spin and angular momentum independently of the electron
distribution between the two parts of the confinement potential.
The electron distribution in the system can be illustrated by the
charge accumulated within the dot. This quantity is calculated as
the integral of the radial probability density from the origin to
the cusp of the confinement potential (cf. Fig. 1). Fig.
\ref{fig2}(a) shows the dependence of the charge accumulated
within the dot as function of the ring oscillator energy for
different barrier thickness, the dot confinement energy $\hbar
\omega_i=6$ meV and equal depth of the dot and ring ($V_0=0$). For
$b=30$ nm the dependence of the charge accumulated in the dot on
the ring confinement energy is almost stepwise and it becomes
smoothened for thinner barriers for which the separation of
electrons between the two parts of the system is less distinct.
The transition of electrons between the ring and the dot can also
be provoked by changing the relative depth of the confinement
potentials for fixed oscillator energies. This is illustrated in
the inset to Fig. \ref{fig2}(a) which shows the charge accumulated
within the dot as a function of $V_0$ for the potential parameters
$\hbar \omega_i=6$ meV, $\hbar \omega_o=20$ meV, $b=30$ nm, i.e.,
corresponding to the central plateau of the solid curve in the
main part of Fig. \ref{fig2}(a).

Fig. \ref{fig2}(b) shows the phase diagram for the spatial
distribution of electrons in the two-electron system in the
absence of the magnetic field for barrier thickness $b=30$ nm.
Borders of regions corresponding to different electron
distributions are marked with solid lines. Above the dotted line
the ground state of a single electron is localized in the dot and
below it -- in the ring. The dotted line can be very well
approximated by $\omega_o=2\omega_i$, which is in agreement with
the approximate formulas for the lowest-energy dot- and ring-
localized states given in the preceding section. In the
noninteracting case this line would divide the regions in which
both of the electrons are localized in the dot or in the well. In
the presence of interaction a third region in which one of the
electrons is localized in the dot and the other in the ring
appears. This region of electron distribution starts slightly
above the dotted line. This results from the fact that the Coulomb
interaction, smallest for both electrons localized in the ring,
stabilizes the ring-confined ground state for larger $\hbar
\omega_o$ than in the noninteracting case. The central region of
the phase diagram for which one electron resides in the dot and
the other in the ring is particularly interesting from the point
of view of potential spin quantum gate applications.\cite{GB}

\begin{figure}[htbp]{\epsfxsize=75mm
                \epsfbox[70 29 491 762]{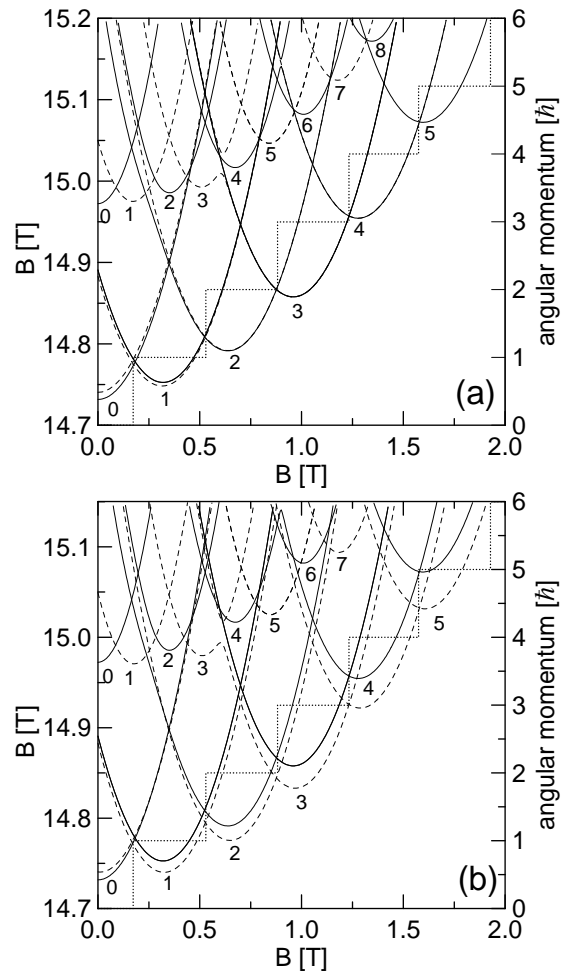}}\newline
%c:\luty\nci2d_latcaoup_harmonic\publishable\TBP\2e\fig3>
\caption{ (a) Two-electron energy spectrum (left scale) for $b=30$
nm, $V_0=0$, $\hbar \omega_i=6$ meV and $\hbar \omega_o=14$ meV
(spin Zeeman effect neglected). Singlets (triplets) plotted with
solid (dashed) lines. Numbers close to extrema of the lines denote
the absolute values of the angular momentum in $\hbar$ units. The
dotted line shows the absolute value of the angular momentum of
the ground state of a {\it single} electron confined in the ring
(right scale).
 \label{fig3} (b) Same as (a)
but with spin Zeeman effect included. Only the lowest energy level
of the split spin triplet is plotted.}
\end{figure}

Let us now look at the magnetic field dependence of the
two-electron energy spectrum for the potential parameters
corresponding to one electron in the dot and the other in the
ring, i.e., for $V_0=0$, $b=30$ nm, $\hbar \omega_i=6$ meV and
$\hbar \omega_o=14$ meV presented in Fig. \ref{fig3}(a). For this
potential the one-electron ground state is localized in the dot.
The angular momentum of the lowest excited ring-confined
one-electron state is plotted with a dotted line (right scale).
Comparison of this line with the ground state energy crossings in
the two-electron spectrum shows that the angular momentum
transitions in the two-electron system are due to the
Aharonov-Bohm effect for the electron confined within the ring.
All the angular momentum of the system is therefore carried by the
ring-confined electron while the electron confined in the dot
remains in the $s$ state. Singlet-triplet splitting of the ground
state [cf. the distance between the dashed and solid lines in Fig.
\ref{fig3}(a)] disappears at larger angular momentum. This effect
can be understood if we look back at Fig. \ref{fig1}(c) showing
that the dot penetration of the ring-localized single-electron
states decrease with their angular momentum.
In Fig. \ref{fig3}(a) above 14.9 meV the energy band corresponding
to both electrons confined within the ring appears. Since in this
band both ring-confined electrons are subject to the Aharonov-Bohm
effect the angular momentum of the lowest state in the band grows
roughly twice\cite{CEPL} as fast as in the ground-state. The
energy levels of even $L$ correspond to spin singlets and of odd
$L$ to triplets. Around 0.6 T we observe an anticrossing of $L=3$
triplets corresponding to one and two electrons in the ring. The
Zeeman effect [cf. Fig. \ref{fig3}(b)] for large $B$ lifts the
ground-state degeneracy with respect to the spin.

\begin{figure}[htbp]{\epsfxsize=75mm
                \epsfbox[21 57 572 576]{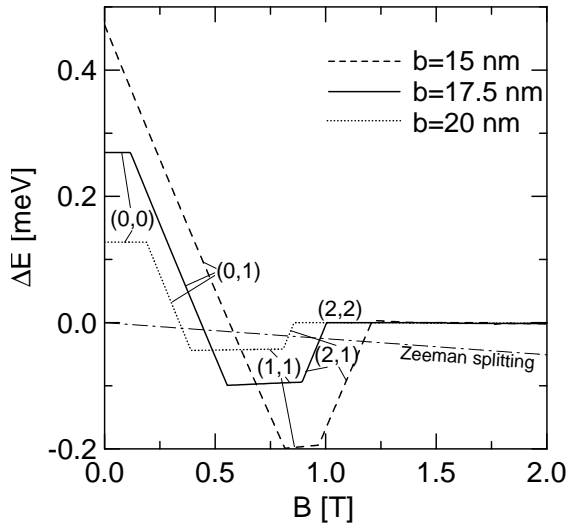}}\newline
%c:\luty\nci2d_latcaoup_harmonic\publishable\TBP\2e\fig4_exchange>
\caption{Exchange energy, i.e., the energy difference of the
lowest triplet and the lowest singlet energy levels for two
electrons and $V_0=0$, $\hbar \omega_i=6$ meV and $\hbar
\omega_o=18$ meV for different values of the barrier width and
spin Zeeman splitting is neglected. The dash-dotted line shows the
Zeeman splitting between states with $S_z=0$ and $\hbar$.}
 \label{fig4}
 \end{figure}

The energy difference between the lowest spin singlet and triplet
states, i.e. the exchange energy\cite{GB} -- an important quantity
for the coupled spin qubit operations is also a very adequate
measure of the strength of the tunnel coupling between the dot and
the ring confined wave functions. Fig. \ref{fig4} shows the
exchange energy (Zeeman energy neglected) for different values of
the barrier thickness for $V_0=0$, $\hbar \omega_i=6$ meV and
$\hbar \omega_o=18$ meV, i.e., for the central point of the
plateau corresponding to one of the electrons localized in the dot
[cf. Fig. \ref{fig2}(a)]. The exchange energy is nearly
independent of magnetic field when the lowest singlet and the
lowest triplet possess the same angular momentum and it distinctly
decreases (grows) with the magnetic field when the $L$ of the
lowest triplet is larger (smaller) than $L$ of the lowest singlet.
When the angular momentum of both singlet and triplet states
exceed $2$ the exchange energy vanishes. The exchange energy is a
piecewise linear function of the magnetic field in contrast to
smooth oscillatory dependence of the exchange interaction on the
magnetic field for side-by-side dots (cf. Fig. 4 of Ref.
[\cite{Harju}]). In side-by-side dots the magnetic field induces a
continuous decrease of the overlap of the wave functions of
electrons confined in different dots. For the dot in the ring
geometry this decrease is discontinuous due to the Aharonov-Bohm
effect for the ring confined electron. Since the Aharonov-Bohm
magnetic period is inversely proportional to the square of the
ring radius one can largely reduce the range of the magnetic field
in which the exchange energy is nonzero by a mere increase of $R$.

\begin{figure}[htbp]{\epsfxsize=75mm
                \epsfbox[27 154 560 665]{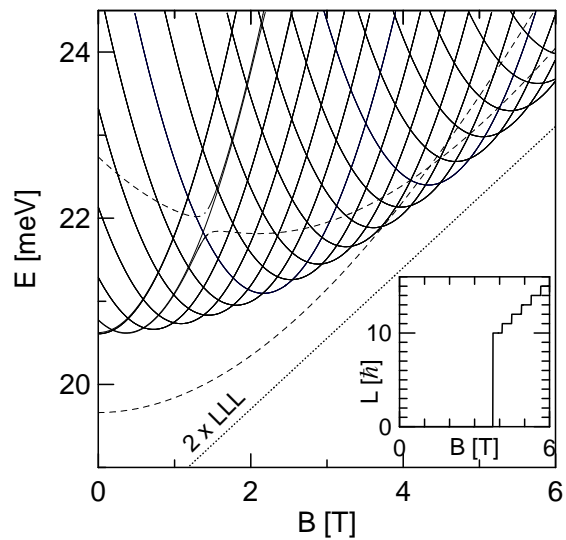}}\newline
%c:\luty\nci2d_latcaoup_harmonic\publishable\TBP\2e\fig4_exchange>
\caption{ Energy spectrum of two electrons for $b=30$ nm, $\hbar
\omega_i=6$ meV and $\hbar \omega_o=26$ meV (spin Zeeman effect
neglected). The energy levels of states in which both (one)
electrons are localized in the dot are plotted with dashed (solid)
lines. \label{30_6_26} The inset shows the ground state angular
momentum. The dotted line corresponds to twice the lowest Landau
level energy.
 }
\end{figure}

The magnetic field can change the distribution of the electrons
between the coupled cavities. Consider the case of $b=30$ nm,
$V_0=0$, $\hbar \omega_i=6$ meV and $\hbar \omega_o=26$ meV. For
these parameters in the absence of the magnetic field both
electrons are localized within the dot [cf. Fig. \ref{fig2}(b)],
but the state corresponding to one electron in the ring is close
in energy. Fig. \ref{30_6_26} shows the magnetic field dependence
of the two-electron energy spectrum for this potential. Energies
of states corresponding to both electrons localized in the dot are
plotted with dashed lines. The lower of these two energy levels is
a spin singlet of $s$ symmetry. The upper dashed line corresponds
to a spin triplet of $p$ symmetry, i.e., to the two-electron
maximum density droplet.\cite{mdd} Spin singlet of $p$-symmetry
with both electrons localized in the dot lies higher in energy
beyond the range presented in this figure. The energy levels
plotted with solid lines correspond to one electron localized in
the dot (in the lowest $s$ state) and the other in the ring. For
$B=1.44$ T an avoided crossings appears for the $L=1$ spin
triplets. For $B=3.74$ T the energy level of the dot localized
state crosses the energy level of the state with $L=10$
corresponding to one electron in the dot and the other in the
ring. Note that below $B=3.74$ T in the ground-state the electrons
are in the singlet state while above this field singlet and
triplet states are nearly degenerate. Decoupling of spins, in the
sense of vanishing exchange energy appears abruptly after crossing
$B=3.74$ T. For $B=4.35$ T a crossing of dot-localized singlet and
triplet states appears. The dotted line shows twice the lowest
Landau energy level. For $B>4$ T the envelope of the energy levels
with one electron in the dot and the other in the ring as well as
the dot localized maximum density droplet run approximately
parallel to the lowest Landau level (dotted line). Fig.
\ref{30_6_26} shows that the magnetic field can change the
electron occupation of the dot and the ring. Generally, for
$V_0=0$ such an effect is not observed for a single electron. The
appearance of this effect for two electrons is due to lowering of
the Coulomb interaction energy when one of the electrons is
transferred from the dot to the ring. Recently,\cite{bs2} it was
shown that in the infinite magnetic field limit the ground-state
electron distribution can be identified with the lowest energy
configuration of classical\cite{Bedanov} point charges. For
$V_0=0$ the lowest-energy classical configuration corresponds to
both electrons localized in the ring. One should therefore expect
that at higher magnetic fields the second electron should also be
transferred to the ring. However, the magnetic fields at which
this effect could appear are beyond the reach of our numerical
calculations.

\begin{figure}[htbp]{\epsfxsize=75mm
                \epsfbox[14 230 576 706]{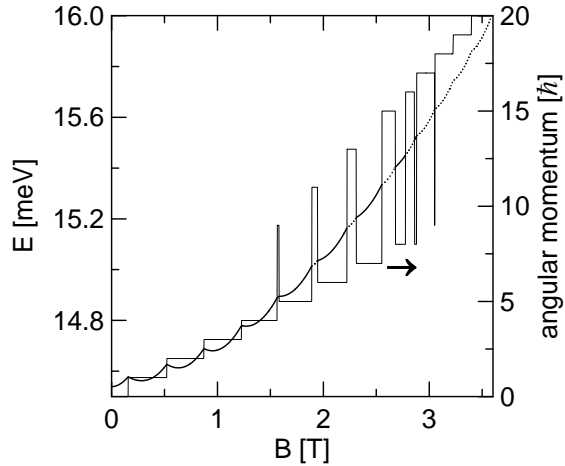}}\newline
%c:\luty\nci2d_latcaoup_harmonic\publishable\TBP\2e\b30\6_1365>
\caption{ Two-electron ground-state energy  (left scale) for
$b=30$ nm, $V_0=0$, $\hbar \omega_i=6$ meV and $\hbar
\omega_o=13.65$ meV (spin Zeeman effect neglected). Energy of
states corresponding to one electron in the dot and the other in
the ring plotted with solid line. Energy of states in which both
the electrons are localized in the ring are plotted by the dotted
curve. The thin solid step-like line gives the total angular
momentum which is referred to the right axis. \label{inout}
 }
\end{figure}

For $V_0=0$, $b=30$ nm, $\hbar \omega_i=6$ meV and $\hbar
\omega_o=13.65$ meV [the left end of the central plateau in the
Fig. \ref{fig2}(b)]  for $B=0$ one of the electrons is localized
in the dot and the other in the ring, but the state with two
electrons localized in the ring is not much higher in energy. Fig.
\ref{inout} shows the ground state energy and the ground-state
angular momentum as function of the magnetic field for this set of
parameters. The state with one electron in the dot remains the
ground state up to 1.6 T. Between $B=1.6$ T and $B=3.1$ T the
state with two electrons in the ring is almost degenerate with the
state with one ring-confined electron and as a consequence the
localization of the ground-state changes several times as the
magnetic field is increased. Ground-state ring-localization
becomes established above 3.1 T. The period of the angular
momentum transitions becomes halved with respect to the low
magnetic fields, for which the ring is occupied by a single
electron.

\begin{figure}[htbp]{\epsfxsize=75mm
                \epsfbox[17 139 588 660]{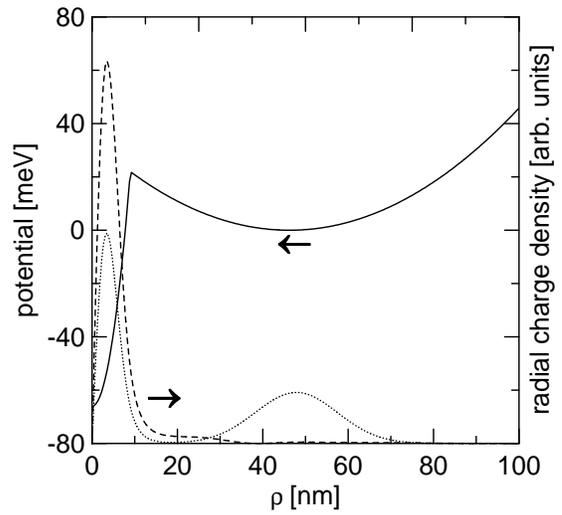}}\newline
%c:\luty\nci2d_latcaoup_harmonic\publishable\TBP\2e\unequal\50_6_20_66>
\caption{Solid line (left scale): the external potential $\hbar
\omega_i=50$ meV and $\hbar \omega_o=6$ meV, $b=20$ nm, $V_0=-66$
meV. The ground state for $B=0$ corresponds to the $L=0$ singlet
with one electron in the dot and the other in the ring. Radial
density of this state plotted with dotted line (right scale). The
first excited $s$ singlet state corresponds to both electrons in
the dot (dashed line).}
 \label{smalldot}
\end{figure}

\begin{figure}[htbp]{\epsfxsize=75mm
                \epsfbox[17 174 562 683]{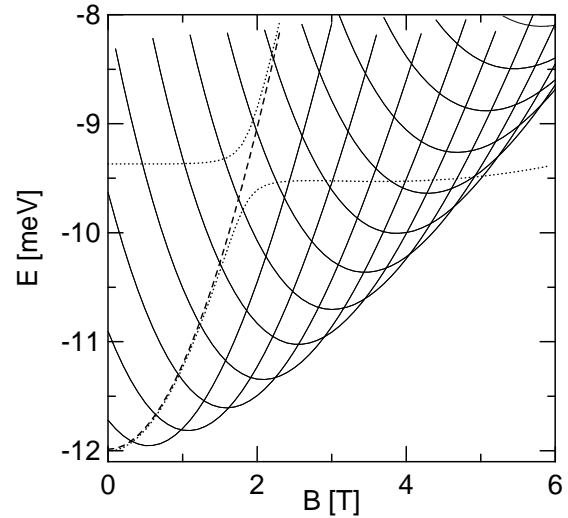}}\newline
%c:\luty\nci2d_latcaoup_harmonic\publishable\TBP\2e\unequal\50_6_20_66>
\caption{Two-electron energy spectrum for the potential parameters
of Fig. \ref{smalldot}. Dotted lines show the energy levels of $s$
singlets. The dashed line corresponds to the $s$ triplet. }
 \label{smalldotbe}
\end{figure}

The magnetic field can induce opposite transitions of the
electrons from the ring to the dot if the dot is small but deep.
Consider the following set of parameters $\hbar \omega_i=50$ meV,
$\hbar \omega_o=6$ meV, $b=20$ nm, and $V_0=-66$ meV. Fig.
\ref{smalldot} shows the confinement potential and the radial
probability density for the lowest two-electron singlet states
with total angular momentum equal 0. For zero magnetic field in
the ground-state one electron is localized in the ring and the
other one in the dot. In the first excited $s$ singlet state both
the electrons reside within the dot. Note that in this case the
ground state is more extended than the excited state  as a
consequence of the electron-electron interaction preventing the
second electron from entering the dot. The magnetic field energy
dependence is displayed in Fig. \ref{smalldotbe}. The magnetic
field has only a small influence on the energy of the singlet with
both electrons localized in the dot. Around $B=2$ T singlets
corresponding to different electron distribution change their
energy order with a pronounced anticrossing. For $B=5.725$ T the
dot-localized singlet becomes the ground state. In this structure
the Aharonov-Bohm oscillations are interrupted by the magnetic
field which removes the second electron from the ring. As a
consequence a giant singlet-triplet energy difference appears for
$B>5.725$ T. This transition appears in spite of the Coulomb
interaction energy which is increased when the second electron is
trapped in the central cavity.

\section{three electrons}

\begin{figure}[htbp]{\epsfxsize=75mm
                \epsfbox[13 77 562 603]{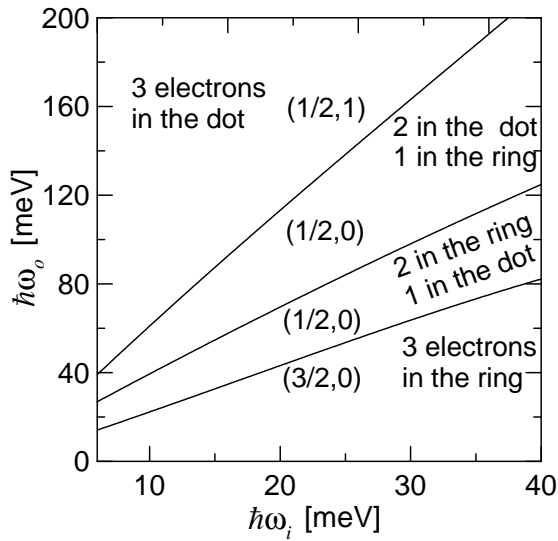}}\newline
%c:\luty\nci2d_latcaoup_harmonic\publishable\TBP\2e\unequal\50_6_20_66>
\caption{Phase diagram for the electron distribution in the
$(\hbar \omega_i$, $\hbar \omega_o)$ plane for $V_0=0$ and $b=30$
nm in the absence of the magnetic field. Solid lines separate
regions of different electron distributions. Numbers denote the
ground-state total spin and total angular momentum quantum number
$(S,L)$. \label{f3e}
 }
\end{figure}

Distribution of electrons in the three electron system without the
magnetic field for $V_0=0$ and $b=30$ nm is plotted in Fig.
\ref{f3e}. Regions of different electron distribution are
separated by the solid lines. For large dot confinement energy,
i.e., $\omega_i>>\omega_o$ all the electrons reside in the ring
and the ground state corresponds to total spin $S=3/2$ and
zero\cite{CEPL} angular momentum. In the single-particle picture
this state corresponds to electrons having parallel spin and
occupying states with angular momentum $l=-1,0$, and $1$ (in
$\hbar$ units). For increasing ring confinement the electrons
enter the dot one by one. In states with two electrons of opposite
spins occupying the dot or the ring (cf. two central regions in
Fig. \ref{f3e}) the ground state corresponds to $S=1/2$ and $L=0$.
When the ring confinement energy is much larger (5 times or more)
than the dot confinement all the electrons occupy the lowest
dot-confined energy levels forming the state of spin $S=1/2$ and
angular momentum $L=1$.

\begin{figure}[htbp]{\epsfxsize=75mm
                \epsfbox[34 143 581 655]{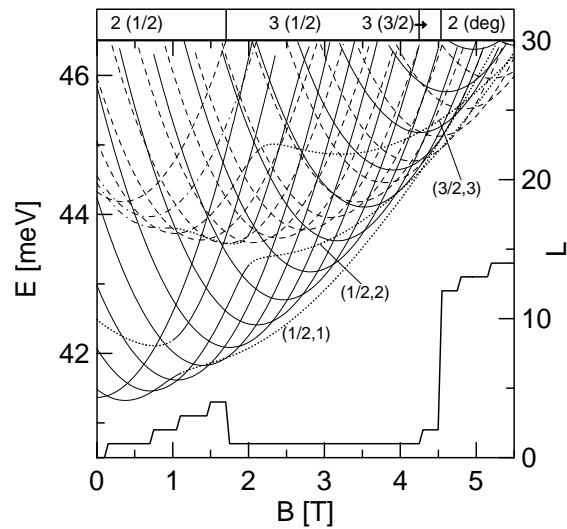}}\newline
%c:\luty\nci2d_latcaoup_harmonic\publishable\TBP\2e\unequal\50_6_20_66>
\caption{ Energy spectrum for $N=3$, $V_0=0$, $b=30$ nm, $\hbar
\omega_i=6$ meV and $\hbar\omega_o=37$ meV. The solid (dashed)
lines show the lowest energy levels with the two dot-confined
electrons electrons with $S=1/2$ and opposite ($S$=3/2 and the
same) spin and one electron in the ring. The states with the two
dot-confined electrons of parallel spins are almost degenerate
with respect to the spin orientation of the electron in the ring.
The only exception is the state with $L=2$. The lower of the
dash-dotted line shows this state for $S=1/2$ and the upper for
$S=3/2$. Dotted lines correspond to all electrons confined in the
dot. Quantum numbers $(L,S)$ of these states are indicated in the
figure. Only non-positive angular momenta are shown. Thick solid
line in the lower part of the figure shows the the ground-state
angular momentum quantum number (right scale). The panel above the
upper axis shows the number of electrons in the dot $n_d$ and $S$
for the ground state in format $n_d(S)$, 'deg' stands for
degeneracy of the $S=1/2$ and $3/2$ states. \label{dodot}
 }
\end{figure}

\begin{figure}[htbp]{\epsfxsize=75mm
                \epsfbox[36 180 573 635]{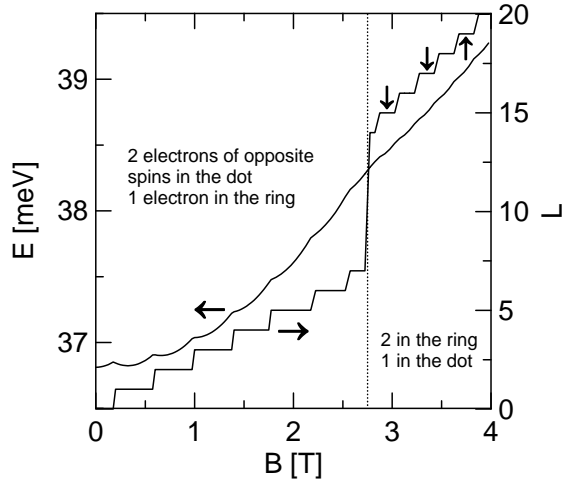}}\newline
%c:\luty\nci2d_latcaoup_harmonic\publishable\TBP\2e\unequal\50_6_20_66>
\caption{Ground-state energy (left scale) and the absolute value
of the ground-state angular momentum (right scale) for $N=3$,
$V_0=0$, $b=30$ nm, $\hbar \omega_i=6$ meV and $\hbar\omega_o=27$
meV. The dotted vertical line marks the magnetic field for which
the electron distribution is changed. The vertical arrows on the
$L$ staircase correspond to triplet state of the ring subsystem.
\label{3efin}
 }
\end{figure}

In the preceding section we showed that for equal depth of the
ring and the dot $(V_0=0$) the electron-electron interaction
triggered the magnetic-field-induced transitions of electrons from
the dot to the ring. We found that in the three-electron system
the magnetic field can also induce the opposite transition from
the ring to the dot. This is illustrated in Fig. \ref{dodot} which
shows the energy spectrum for $V_0=0$, $b=30$ nm, $\hbar
\omega_i=6$ meV and $\hbar \omega_o=37$ meV. Solid lines in Fig.
\ref{dodot} show the energy levels corresponding to two electrons
of opposite spins in the dot and one electron confined in the
ring. All these states have $S=1/2$. Dashed lines correspond to
spin-polarized states with $S=3/2$ in which the two dot-confined
electrons occupy the $1s$ and $1p$ states. Energy levels
corresponding to three electrons localized in the dot are shown by
dotted curves. Quantum numbers of the dot-confined states are
given in the figure. Thick solid step-like line at the bottom of
the figure shows the absolute value of the ground-state angular
momentum which is referred to the right axis.  At $B=0$ the energy
of the state in which all three electrons are localized in the dot
with $L=1$ is 1 meV higher in energy (cf. dotted line above 42
meV) above the ground state with two electrons in the dot and one
in the ring. This energy level decreases initially with increasing
magnetic field due to the interaction of the magnetic field with
the magnetic momentum of the $p$ electron. This decrease results
in an anticrossing of the $L=1$ energy levels corresponding to two
and three dot-confined electrons around $B=1$ T. Another
consequence of this anticrossing is a visibly increased region of
$L=1$ ground state stability between 0.15 and 0.7 T. Subsequently
for $B=1.7$ T the state with three electrons in the dot and $L=1$
becomes the ground state. The transition of the third electron
from the ring to the dot happens in spite of the electron-electron
interaction which is not strong enough to prevent it.

For $B=0$ the energy of the lowest spin polarized state (cf.
dashed lines) with $L=1$ is equal to about 44.25 meV. In this
state the two electrons confined in the dot have the same spin and
occupy $1s$ and $1p$ energy levels, while the ring-confined
electron occupy the orbital with $l=0$. Note that level crossings
appear at the same magnetic fields as in the lower branch with
$S=1/2$ where two electrons are in the 1s orbital confined in the
dot (cf. solid lines in the Fig. 13). Angular momentum quantum
number of these states is equal to the ring confined electron plus
1 -- the angular momentum of the dot-confined subsystem. For the
adopted large barrier thickness $b=30$ nm the states of this band
with $S=3/2$ are almost degenerate with $S=1/2$ states (omitted in
the figure), i.e., the energy of the system is not influenced by
the orientation of the spin of the ring-confined electron. The
only exception appears for the $L=2$ state. The lower (upper)
dashed-dotted line shows the energy of the state with $L=2$ and
$S=1/2$ ($3/2$). The reason of the lifted degeneracy is the fact
that the energy of the state with $L=2$ and $S=1/2$ is pushed
downward by the anticrossing with the dot-confined state of the
same quantum numbers, similarly as the above discussed $L=1$,
$S=1/2$ energy level in the lower part of the spectrum.

The angular momentum of the ground state with three-electrons
confined in the dot changes from $0$ to $1$ at $B=4.3$ T (cf. the
crossing of the dotted lines). Above 4.5 T the ground state
corresponds again to two electrons in the dot and one in the ring
like for $B=0$ T, but now the dot-confined subsystem is
spin-polarized (cf. the crossing of the dashed and dotted lines).

Fig. \ref{3efin} shows the ground-state energy for the same
parameters as studied in Fig. \ref{dodot} but with the ring
confinement energy reduced from 37 to 27 meV. At $B=0$ the ground
state still corresponds to two electrons in the dot and one in the
ring, but the state with two ring-confined electrons is higher in
energy by less than 1 meV. The envelope of the lowest energy level
with one electron in the ring grows with the magnetic field faster
than the envelope of the energy levels with two ring-confined
electrons which results in the change of the ground-state electron
distribution at $B=2.75$ T. The dotted line in Fig. \ref{dodot}
marks the change in the electron distribution. At left of this
line the ground state has $S=1/2$, the two dot-confined electrons
are in the spin singlet and the spin of the ring confined electron
is arbitrary. At the right of this line the spin-configuration of
the ring-confined subsystem oscillates from singlet (for even $L$)
to triplet (for odd $L$) [cf. also the branch of ring confined
electrons in Fig. 5(a)]. The magnetic fields for which the ring
subsystem is in the triplet state have been marked by vertical
arrows on the angular momentum staircase. The states with spin
singlet of the ring subsystem correspond to $S=1/2$. Since the
angular momenta of both ring-confined electrons exceed 6 the
tunnel coupling between the dot and ring wave functions is
negligible [cf. Fig. 8] and the spin of the dot confined electron
does not influence the energy. Therefore, the states with triplet
configuration of spins in the ring subsystem correspond to $S=1/2$
and $S=3/2$ degeneracy.

The envelope of the lowest $N=3$ energy levels with three
electrons and two electrons in the ring run almost parallel to
each other as function of the external field. One should
expect\cite{bs2} that for equal depth of the ring and the dot at
very large magnetic field the three-electron ground state
corresponds to electrons forming an equilateral triangle in the
ring, but in the studied magnetic field range we did not observe a
distinct transition of the last electron from the dot to the ring.

\section{Summary and Conclusions}
We have considered a quantum dot inside a quantum ring -- a unique
example of lateral coupling realized with conservation of circular
symmetry of the confinement potential. A simple model for the
potential was used. The model assumes parabolic confinement in
both the dot and the ring so approximate formulas can be given for
the lowest-energy single-electron dot- and ring- confined states.
One-, two- and three-electron systems were studied using the exact
diagonalization approach. We have investigated the distribution of
electrons between the dot and the ring. This distribution depends
not only on the parameters of the confinement potential but it can
also be altered by an external magnetic field, which therefore can
be used as a driving force to transfer the electrons from the dot
to the ring or {\it vice versa}. The passage of an electron from
the dot to the ring should be detectable by a change of the
Aharonov-Bohm magnetic period. The present model allows also for
simulation of the magnetic field induced electron trapping in
local potential cavities. We have studied the exchange energy in
the two electron system with one electron confined in the dot and
the other in the ring. Due to the angular momentum transitions
resulting from the Aharonov-Bohm effect for the ring-confined
electron, the singlet-triplet splitting exhibits a piecewise
linear dependence on the external magnetic field. This should be a
more elegant method for the control of the spin-spin coupling than
the smooth oscillatory dependence predicted for side-by-side
coupled dots.\cite{Harju}

{\bf Acknowledgments} This paper has been partly supported by the
Polish Ministry of Scientific Research and Information Technology
in the framework of the solicited grant PBZ-MIN-008/P03/2003, the
Flemish Science Foundation (FWO-Vl), the Concerted Action
programme (IUAP) and the University of Antwerpen (VIS and GOA).
One of us (BS) is supported by the Foundation for Polish Science
(FNP).
\newline

\end{document}